# The State of the Art Forensic Techniques in Mobile Cloud Environment: A Survey, Challenges and Current Trends

Muhammad Faheem
School of Computer Science and informatics
University College Dublin
Dublin, Ireland
Muhammad.faheem@ucdconnect.ie

M-Tahar Kechadi
School of Computer Science and informatics
University College Dublin
Dublin, Ireland
tahar.kechadi@ucd.ie

Nhien-An Le-Khac
School of Computer Science and informatics
University College Dublin
Dublin, Ireland
an.lekhac@ucd.ie

**Abstract:**

Smartphones have become popular in recent days due to the accessibility of a wide range of applications. These sophisticated applications demand more computing resources in a resource constraint smartphone. Cloud computing is the motivating factor for the progress of these applications. The emerging mobile cloud computing introduces a new architecture to offload smartphone and utilize cloud computing technology to solve resource requirements. The popularity of mobile cloud computing is an opportunity for misuse and unlawful activities. Therefore, it is a challenging platform for digital forensic investigations due to the non-availability of methodologies, tools and techniques. The aim of this work is to analyze the forensic tools and methodologies for crime investigation in a mobile cloud platform as it poses challenges in proving the evidence. The advancement of forensic tools and methodologies are much slower than the current technology development in mobile cloud computing. Thus, forces the available tools, and techniques become increasingly obsolete. Therefore, it opens up the door for the new forensic tools and techniques to cope up with recent developments. Hence, this work presents a detailed survey of forensic methodology and corresponding issues in a mobile device, cloud environment, and mobile cloud applications. It mainly focuses on digital forensic issues related to mobile cloud applications and also analyze the scope, challenges and opportunities. Finally, this work reviewed the forensic procedures of two cloud storage services used for mobile cloud applications such as Dropbox and SkyDrive.

### 1.0 Introduction to mobile cloud computing

Mobile cloud environment utilizes unlimited resources offered by a cloud computing in a resource constraint mobile environment (Liu, Jain, Hu, Zhao, & Zhang, 2009). Mobile Cloud computing is a combination of two new emerging information technology worlds. The motive of the mobile cloud computing concept is to make use of the computing power of the cloud environment and make it available to the mobile devices in order to solve the challenges in a mobile environment. In recently developed mobile cloud architecture, mobile devices can access cloud services either through ad-hoc mobile network or access points (Khan, Othman, Madani, & Khan, 2013). Smartphones are enhanced with powerful hardware in recent years. However, device energy, network connectivity, bandwidth utilization, and mobility are the major constraints for smartphones. The usage of Smartphones has since skyrocketed as it offers several attractive facilities such as surfing the Internet, checking e-mails, video-conferencing, accessing social websites, road navigation, and editing documents. Blackberry, iPhone, and Android Smartphones are most popular in the mobile market. Factors like device's weight, battery life, and heat dissipation limits the applications that can run on these Smartphones. However, the rapidly evolving technology of cloud computing could assist Smartphones to manage these critical factors, by running and storing applications in the cloud. Moreover, mobile cloud computing assists mobile devices to overcome the limitations in terms of data storage, bandwidth, heterogeneity, scalability, availability, reliability, and privacy (Dinh, Lee, Niyato, & Wang, 2013).

Cloud computing is a shared collection of configurable networked resources that can be reconfigured in a short time with minimal effort. In the cloud environment, the cloud service providers maintain data centers worldwide (global cloud exchanges) to guarantee service availability and cost-effectiveness (Buyyaa, Yeoa, Venugopala, Broberg, & Brandic, 2009) and the cloud offers vast storage and facilities to its users. From the cloud user's point of view, it would be an attractive feature for a start-up organization and fabrication line in case of a hardware startup organization (Michael Armbrust, et al., 2009). An essential component of the cloud computing environment, characteristics, service and deploy models are described in (Lovell, 2011) . Mobile devices are capable of performing a collection of functions that ranges from a simple voice call to the complex functions of a personal computer. The imperative characteristics of mobile phones include operating systems, memory considerations, identity module characteristics, and communication protocols. A detailed description and architecture of all mobile device's OSs found in (Yates, 2010) (Casadei, Savoldi, & Gubian, 2006).

With the rise in demand, several threats grow continuously, and there are only limited security features currently available for the mobile device and the platform. Cyber criminals and illegal organizations are likely to take advantage of any emerging technology to commit crimes. Rapidly growing mobile cloud computing technology can be used to broadcast terrorist ideology, share information, and facilitate communication for attacking someone's digital information. Thus, the mobile cloud forensics play an important role in performing forensics in recently developed Smartphone applications that include bump payments, mobile credit billing, and location aware commerce. Mobile cloud forensics deals all illegal activities that utilize cloud to implement them. The mobile cloud forensics established that investigates integration of the virus into cloud applications, utilization of mobile cloud by terrorists, and illegal activities implemented in cloud storage applications such as Dropbox and SkyDrive. In recent days, forensic tools are capable of recovering a localized image in which the virtual machine images move from a cloud-computing platform to the local forensic environment for performing forensics in the virtualization platform.

**1.1 Potential applications of mobile cloud computing**

Applications of mobile cloud are successful in five categories such as productivity, utilities, social networking, games and search (Research:, 2009). The application in the productivity category is most successful among these categories that mainly focus on enterprise-based applications such as data sharing, customer relation management, scheduling, multi-tasking, and merchant services. Moreover, applications of mobile cloud computing include image processing, natural language processing, GPS and Internet access sharing, applications related to sensor data, querying, and crowd computing (Fernando, Loke, & Rahayu, 2013) (Wang, Chen, & Wang, 2014). Apart from these applications, there are other categories of mobile cloud computing applications as listed in the table 1. Among the three commonly used Smartphones, Android mobile applications, introduces the highest security risks of about 43%, followed by Apple iOS (36%) and Blackberry (22%) (Research, 2012). These applications create lots of opportunities and requirements for forensic techniques in mobile cloud applications.

| Application Domain | Mobile Cloud Applications |
|---|---|
| Mobile commerce | Financial applications (mobile transaction, payment, and mobile ticketing), advertising (custom made advertisements), and shopping |
| Mobile learning | Plantation pathfinder, cornucopia, and education tool |
| Mobile health care | Health-aware mobile devices to alert healthcare |
| Mobile Gaming | Graphic rendering |
| Social-Networking applications | Facebook, Twitter, MySpace, LinkedIn, Whatsapp, Viber, Hike, WeChat, Kik messenger and so on |
| Storage services | Dropbox, Sugarsync, Evernote, and SkyDrive |
| Real-time practical applications | Google map, video conferencing, document and image editing |
| Assistive applications | Mobile devices with pedestrian crossing guidance, Mobile currency reader for blind and visually impaired |
| Crowd sourcing | Crowd computing |
| Collecting sensing | Traffic/environment monitoring, and health care |

**Table 1: Application of Mobile Cloud Computing**

**1.2 Recent developments in mobile and its impact on mobile cloud computing**

As the result of the popularity of Smartphones, sales reached 472 million units that are same as 58% of all mobile device sales in 2010 (Goasduff & Pettey, 2012). A recent report says that the number of Android and iOS OSs increased rapidly from 38- 84 million between 2011 and 2012 (Nielsen., 2012) It also added that the average number of applications on a device increased from 32 – 41 and users spend equal time on Smartphone applications that of the web. Moreover, the number of Smartphone sales achieved a record of 207.7 million units in 2012 i.e. 38.3% higher than the previous year (Meulen & J. Rivera, 2013) . International Data Corporation (IDC) has reported that companies selling android devices are gaining more profit in sales compared to other Smartphones. In the latest survey, IDC announced that the worldwide Smartphone market raised 25.3% in the second quarter of 2014, it is a new quarterly record of 301.3 million shipments (IDC).

Most of the today's mobile application exploits the cloud environment. Facebook, Gmail, Whatsapp, and Skype are good instances. A recent survey conducted by MarketsandMarkets estimated that the global mobile cloud market will grow from $9.43 billion in 2014 to $46.90 billion by 2019, rising up 37.8% (counton2) . The high popularity of technology explores the opportunity for misuse (Turnbull & & Slay, 2008). It is likely expected that 90% of companies may support corporate applications for personal Smartphones by 2014 (NetQin, 2011) . It facilitated the users to maintain the corporate and personal information on the same mobile device. Organizations have less control over the misuse of a Smartphone, and it is difficult to trace the misuse. For instance, there is a possibility for uploading a confidential file to Facebook purposely. Thus, it is mandatory that the forensic techniques for mobile cloud applications should grow along with the increased usage and popularity of mobile devices.

**2.0 Digital forensics in mobile cloud computing**
Digital forensics involves in searching the legal evidence from digital computers and storage media. Several tools are readily available for digital forensics investigation that includes open source tools. Though traditional forensic tools examine the data on storage media effectively, recently developed forensic tools consume a long time and large computational power for evidence analysis. Moreover, the influencing factors like a wide range of operating systems and file formats increases the complexity of data examination and the cost of tool development. In a mobile cloud environment, the applications are mainly implemented outside the mobile device and runs in the cloud environment. It is vital to extracting the forensic evidence from the cloud and the third party application providers in addition to that of traditional mobile devices. Mobile forensic examination based on cloud computing facilitates collaborative forensic investigation to be conducted anywhere at any time (Lee & Dowon Hong, 2011). In the future, the current works on digital forensics can focus on the developing standardized and advanced techniques for data interpretation and forensic processing (Simson L. Garfinkel, 2010). The approach in (Sibiya, Venter, & Thomas Fogwill, 2012) proposes a live digital forensic framework for cloud environments. The forensic cloud in (Lee & Dowon Hong, 2011) suggested the investigators to concentrate more on enhancing the investigation process rather than concentrating on technology used in the investigation process. Moreover, it has recommended to speed-up the investigation process using hacking, cracking, and analyzing.

**2.1 The significance of forensics in mobile cloud computing**

The following points posit the motivation for researching digital forensics in mobile cloud computing. Cloud computing is being widely deployed in all IT domains. There are high demands in developing Smartphone and tablet core applications in IT and business fields. Cloud computing is emerging as the dominant information technology that allows mobile applications to operate and communicate. Criminals and terrorists can exploit the massive storage capacity of cloud to perform a range of illegal activities (Zhu, 2011). Therefore, it is essential to extract the data stored or accessed in the cloud from cloud service providers. Therefore, the hefty research investment required for healthy development of tools and technologies to cope with the current pace of technology development in mobile cloud computing. The advantages of mobile cloud computing can be effectively utilized for illegal activity. Thus, law enforcement needs an effective forensic service on mobile cloud computing.

**3.0 Forensic challenges in the mobile cloud computing**

**3.1 Forensic challenges in applications of mobile cloud computing**

This section highlights the forensic issues related to applications of mobile cloud computing.

- The wide range of mobile cloud applications, heterogeneity in mobile cloud architecture and application off-loading techniques, heterogeneity in mobile devices, and heterogeneity in the cloud and storage are the major factors which largely increase the complexity, and delay. It is difficult to obtain and process data in a forensically sound manner.

- The forensic analysis in applications of mobile cloud computing lacks speed. Current forensic tools and technologies require improving data examination speed as it involves a vast amount of digital data.

- There are several mobile cloud applications each of them varies in OSs and supported file formats. As these factors are incessantly changing, there is an increase in the cost of maintenance and tool development. The forensic investigation compulsorily requires examination of all files in different formats and moreover, it is responsible for handling the OS and file system of each device.

- The mobile cloud computing is completely decentralized architecture and elastic in nature. Mobile devices establish wireless connectivity with the remote cloud server. Thus, forensic tools do not cope up with the computing environment. Computers are the component of the cloud architecture and inter-operate within the network without the client's awareness. Moreover, the unauthorized clients can easily seize the client's IDs and passwords due to the open nature of the cloud (Aminnezhad, Dehghantanha, Abdullah, & ohsen Damshenas, 2013).

- Investigations in the mobile cloud application become complex as the operational environment includes several thousands of virtual machines, multiple servers, and numerous cloud clients. In cloud forensics, the physical inaccessibility of servers introduces disruptive issues for investigators (D. Birk, 2011).

- Challenges in the legal process increasingly complicate the scope of the forensic investigation process in mobile cloud applications.

- Lack of standards in the investigation process, tools and techniques.

**3.2 Limitations in the currently available forensic tools for the mobile cloud platform**

Most of the commercial forensic tools can extract the forensic rich data from the traditional mobile phone data, but still there is no well-developed forensic tool for extracting forensic data from third-party applications. However, few commercial tools support the extraction of forensic data from particular third-party application providers. Some of the commonly used mobile forensic extraction tools are XRY, Cellebrite, Oxygen, Zdziarski, and FTK. Most of the existing forensic tools do not support forensics for obstructed or physically damaged mobile devices. Oxygen facilitates the extraction of data from Skype and WiFi connections. Moreover, the developers of commercial forensic tools such as XRY and Oxygen are still working to extract more potential data from the third party application and the device's memory (Zhu, 2011). The recently developed XRY v5.5 extracts Viber chats and call logs that could not be extracted using XRY v5.2. XRY and Oxygen forensic suite successfully perform Android physical extraction. However, XRY can extract only limited data from a Motorola Milestone.

**4.0 Forensic cycle of mobile cloud environment**

The basic steps involved in forensic investigation are discussed in (Kent, Chevalier, Grance, & Dang, 2006). It is important to note that the forensic steps in a mobile device and cloud environment are the same. Forensic analysis on the mobile device is conducted before analyzing the cloud environment. This results in iteration of forensic steps in a mobile device and cloud environment. The following are the general phases of the forensic cycle: identification and preservation, collection, examination and analysis, and presentation.

| Forensic Phases | Mobile Cloud Computing | |
|---|---|---|
| | **Mobile Device** | **Cloud Computing** |
| Identification and Preservation | Identifies the sources of evidence required for performing the digital forensic investigation of the mobile device such as laptops, and Smartphones | Identifies the cloud service provider related to the case and data stored in the cloud storage owned by the corresponding cloud service provider |

| | Collection | Retrieves the actual data from the mobile device. | Retrieves the actual data from the cloud |
|---|---|---|---|
| | Examination and Analysis | Analyzes and discovers whether the physical device used the cloud computing | |
| | Presentation | Document the evidences identified and collected. It maintains a record of all activities and observations that clarifies the outcome of tests and examinations and describes the inferences obtained from the data | |

**Table 2: Forensic cycle**

**4.1 Issues in mobile cloud forensic cycle**

- **Identification and preservation:** The identification of data corresponding to a particular suspect requires more research in a cloud environment. The distributed nature of cloud data causes jurisdictional issues that create a delay in accessing the data. It prevents timely identification and preservation of evidence. The preservation of volatile data on the mobile device and the cloud is a challenging task. It is still difficult to ensure the integrity of the data during the preservation of evidence.

- **Collection:** There is no standard technique to collect deleted data from the volatile and non-volatile memory of the device and also from the cloud storage.

- **Examination and analysis:** The examination and analysis involve in dealing with a vast amount of data. Thus, the examination of evidence relevant to the case for all other data poses more challenges. The selection of appropriate tools for analyzing the evidence and the availability of selected tools is a great challenge. Moreover, the analysis of encrypted data poses more difficulties.

- **Presentation:** The incomplete or inconsistent report can lessen the correctness of the evidence gathered. It is necessary to eliminate the inconsistencies carefully if any, in the document before the presentation.

## 5.0 Mobile device forensics
## 5.1 Digital forensics in mobile devices

Mobile device forensics is a part of digital forensics. Mobile forensics deals with the process of gathering evidence from a mobile device in a forensically sound conditions using well-developed tools and techniques. Mobile forensics involves in five phases: Preservation, Acquisition, Examination, Analysis, and Reporting. Preservation is the process of seizing and securing suspected mobile devices without modifying the contents of data stored on the devices. Acquisition is the process of retrieving information from a mobile device and peripheral equipment. Examination and analysis apply forensic tools to discover potential evidence of the mobile device including hidden or obscured evidences. The investigation of the case comes to an end with reporting that maintains a record of all conclusions drawn from the previous phases (Raghav & Saxena, 2009).

## 5.2 Forensic challenges in mobile devices

- Diverse models of mobile phones have been used in world-wide. However, there is no standardized form for storing the mobile data.
- Mobile forensics lacks standardized extraction techniques. Physical acquisition techniques gather more desired data. However, these techniques are expensive and time consuming, and require high technical expertise. Logical acquisition techniques are faster, but, it fails to guarantee the data integrity. It requires mobile devices to be powered ON to perform the acquisition.
- The absence of write blocking mechanism makes the extracted data less forensically sound.

**5.3 Data acquisition and analysis in mobile device**

A software tool acquires a physical copy of the internal flash memory of the mobile device that operates on Symbian OS and stores the acquired data on a removable disk such as memory card (Me, 2008). It is designed as forensically sound software tools that have a built-in Application Programming Interface (API) to retrieve data from the flash memory in a read-only mode. The mobile device needs to be turned-off prior conducting data acquisition. Insert a new memory card after extracting data from the SIM and Memory cards. The installation of a seized application facilitates direct execution and does not require a physical installation of the mobile device. One can remove the memory card after copying the internal memory of the mobile device onto it. The software tools used in this methodology of data extraction do not support other OSs such as Android and iPhone OS. However, there is a need to develop different software tools to conduct data acquisition successfully.

**5.4 Android forensics**

Android is an OS developed by Google for mobile devices, especially for smartphones and tablets. The architecture of android has four levels, such as the Linux Kernel, Libraries and Android Runtime, Application framework, and Applications. The major application repository of the android mobile device is Google Play Store.

The list of methods for performing forensics investigations in android-based mobile devices found in (Lessard & Kessler, 2009). In android mobile devices, the SQLite databases contain most of the forensic-rich information. The forensic investigation physically extracts the information stored in the memory card using AcessData FTK Imager v2.5.1. Rooting can obtain the physical image of the phone memory. It uses AccessData Forensic Tool Kit (FTK) v1.81 for performing rooting. It generates a logical DD image for conducting the logical examination and analysis using Cellbrite. Logical examination also requires root access. However, the logical image of all the deleted SMS, call logs, and phonebook information could not be extracted. Physical extraction can obtain all this information but consumes a long time for data analysis. The pros and cons of physical and logical extraction enable the examiner to select the type of extraction based on each forensic scenario.

In the formerly developed acquisition methods for Android OS, the investigator must have basic privileges for performing the analysis on the phone. Therefore, it motivated the researchers to develop booting techniques into the recovery mode, if the device had been switched-off (Vidas, Zhang, & Christin, 2011) . Write blocker prevents altering the state of the device. It provides a detailed description of commonly used logical acquisition methods such as adb, AFLogical tool, and Cellebrite UFED in addition to JTAG and bootloader acquisition techniques (physical). The recovery mode ensures data integrity during the preservation (Son, Lee, D. Kim, Lee, & Lee, 2010).

The most frequently used file systems on the devices using the Android OS are YAFFS and YAFFS2. The experimental research in (Quick & Alzaabi, Forensic analysis of the android file system yaffs2, 2011) performs logical and physical acquisition using adb pull, NANDump, xRecovery and Yaffs2utils tools on a Sony Xperia 10i device model. The logical acquisition could not retrieve the full size of the file system, whereas the physical acquisition retrieved content from the flash memory. The use of hex viewer WinHex identifies the header of the retrieved files. NANDump can retrieve a complete copy of the internal NAND memory. The Linux oriented forensic tools are not applicable for Android OS (Sylve, Case, Marziale, & Richard, 2012).

An acquisition method considered the use of wireless networking devices such as WiFi and Bluetooth on four Android OS based devices. There was no big difference in the results even if the devices were of different OS versions (Andriotis, Oikonomou, & Tryfonas, 2012) . Live forensic approaches (Guido, et al., 2013) observe the illegal or criminal activities on Android devices. The architecture comprises of five components in which each component is responsible for detecting modifications in specific modules of OS such as bootloader, recovery, file system, deleted files or folders, and APK files. It fails to consider false positives and do not detect any of the deleted files.

DroidWatch (Grover, 2013) is an android application that continuously tracks the events and data flow on the device and forwards the information to a Web Server. Content providers allow accessing the data residing on other applications, and Droid Watch broadcast receivers track events and content observers tracks database changes.

**5.5 iOS forensics**

Apple has a dedicated cryptographic chip that enables hardware based encryption, and integrated it with all iOS devices. This feature of incorporating cryptography into the OS is termed as Data Protection. Data protection allows encryption or decryption of any file or part of the file using a separate key. The innovations presented in (Zdziarski & Media, 2008) achieved a milestone of implementing a physical acquisition technique for iOS. It concentrates on changing the amount of data in the system partition and fails to consider user data partitioning. It uses a recovery toolkit that contained the essential software for retrieving a bitwise copy of the memory. Several services and techniques tamper to obtain knowledge on the subject, including the manufacturer's back door services (Zdziarski J. , 2014).

**5.6 Application acquisition on mobile device**

The technique proposed in (Levinson, Stackpole, & Johnson, 2011) conducts forensic investigations on the third-party application installed on Apple mobile device. It introduced the type of evidence that can be discovered from the iPhone flash memory and methods to read those evidences. Third party applications, undoubtedly contain forensic-rich information. However, commercial forensic tools only read phone data on Apple mobile devices. Most of the mobile forensic work concentrates on recovering typical mobile data such as contact details, and text and voicemail messages. However, there also remain other forensic artifacts related to third party applications running on mobile devices. It acquires data about user accounts, timestamps, Geo locational references, additional contact information, native files, and various types of media files from 8 third party applications. These applications are Twitter, Facebook, Skype, FourSquare, BrightKite, Where.com, iBooks, and Yelp. The acquired data stored in a plaintext format.

The physical and logical acquisition of WeChat on iPhone acquires data from internal flash memory and logical storage files respectively (Gao & Zhang, 2013). Chip-off extraction performs physical acquisition, and MOBILedit performs logical acquisition. The folders of WeChat are also analyzed such as MM.SQLite files, audio, video and picture files. Though, it extracts digital evidences in iPhone; it fails to address forensic investigation of WeChat application on the cloud environment.

Forensic analysis and recovery of events performed via social networking applications on various Smartphones using different OSs have addressed in (Mutawa, Baggili, & Marrington, 2012). The applications on Facebook, Twitter, and MySpace are experimented with Blackberry, iPhone, and Android devices. There are no artifacts that could be extracted from the Blackberry device. However, it recovers only the basic information such as username, password, timestamp, and viewed and uploaded images on iPhone and Android devices.

Skype is the most famous VoIP application and data stored in Skype act as valuable evidence for digital investigations. Though Skype assures secure communication over the Internet, information of Skype calls and chats are recoverable in android devices (Al-Saleh & Forihat, 2013) . Forensics service retrieves this information from RAM and NAND flash memories of

Android devices. The investigators can acquire more information as a result of E-mail forensics through Visualize Association Inside Emails (VAIE) system (Meng, Wu, Yang, & Yu, 2009). Two layout models such as the spring force model and radial tree model enable the investigator to visualize the forensic results.

## 6.0 Cloud Computing Forensics
### 6.1 Digital forensics in cloud computing

Digital forensics applies the concept of computer science to retrieve electronic evidence for reporting in a court of law. Cloud forensics is an interdisciplinary of cloud computing and digital forensics. There are no forensic capabilities between the cloud service providers and customers that could assist investigations of illegal activities in the cloud (Ruan, Carthy, & T. Kechadi M. Crosbie, 2011). (Ruan K. , Carthy, Kechadi, & Baggili, 2013) and (Aminnezhad, Dehghantanha, Abdullah, & ohsen Damshenas, 2013)[24] discusses the fundamental challenges of cloud forensics and (Dykstra & A. Sherman, Understanding issues in cloud forensics: Two hypothetical case studies, 2011) describe the cloud forensics challenges using two hypothetical cloud crimes. The storage services of cloud include Dropbox, box, and SugarSync. Amazon Cloud Drive facilitates users to transfer the file involving several computers with or without the use of a stand alone application installed on the user's computer (Hale, 2013). Smartphones can preserve the data stored on these storage services (Grispos, Glisson, & Storer, 2013) .

The attractive feature of cloud computing is the centralized data that support forensic readiness, especially in a scenario where it requires a quicker response to incidents or crime (Reilly, Wren, & Berry, 2011). The storage capacity in petabytes and enormous computational resources in the cloud facilitates the forensic investigator to take great advantage. Moreover, it allows the authentication of disk images using inbuilt hash authentication. In the aspect of forensic investigation, the virtualization concept of cloud can be treated both as an advantage and a drawback. Virtualization permits users to share the same resources, and it allows virtualization of several resources such as platform, software, and infrastructure. In contrast, data considered as evidences involve bit-by-bit duplication using suitable software. It takes a long time for the investigators to remove this duplication. There are potential risks in ensuring the privacy of the data in cloud computing (Daryabar, Dehghantanha, Udzir, & Mohd, 2013).

The impact of the nature of cloud in digital forensics could be both harder and easier (Ruan K., Carthy, Kechadi, & Baggili, 2013). Some of the features of cloud make the digital forensics difficult, and some cloud features make the digital forensics easier. Table 3 describes the impact of cloud computing in digital forensics in terms of harder aspects and easier aspects.

| Harder aspects | Easier aspects |
|---|---|
| Reduced access to widely distributed physical machines and storage systems | On demand computing power, supports elasticity, and cheaper |
| Not aware of the exact physical location of the data | Supports scalability in auditing, logging, reporting, and imaging |
| Separation of data accessed by the suspect from multi-users of cloud | Dependency of cloud service provider |
| Legal issues | Suspect cannot easily destroy the evidence in cloud as it is stored in multiple locations |
| Lack of proprietary interface | Supports hashing and imaging |

Table 3: Harder and easier aspects of cloud on digital forensics

## 6.2 Forensic challenges in cloud computing

- There are several attractive features of cloud computing, but still the impact to standard forensics is still unproven, as cloud forensics is in the preliminary stage. The current methods of investigating cloud services are not feasible for efficient execution due to geographical, privacy and legal constraints.

- Some of the research works emphasized the limitation of current cloud forensic tools (Ruan, Carthy, & T. Kechadi M. Crosbie, 2011) (Reilly, Wren, & Berry, 2011) (Grispos, Storer, & Glisson). Development of tools and techniques for performing investigations in virtualized environment, especially on the hypervisor level is still in the infant stage. The cloud is involved in cross-platform development, and lack of standardized infrastructure makes it difficult to develop the tools for forensic investigation. There is a requirement of forensic aware tools for the cloud service provider and the users to gather forensic evidence (Ruan, Carthy, & T. Kechadi M. Crosbie, 2011).

- Moreover, investigators rely on the cloud service providers to gather evidence from a cloud. The cloud service provider does not guarantee the correctness of the evidence it provides. Apart from these factors, issues associated with virtualization, large scale data processing, and growth of mobile devices makes the investigation process much harder.

- According to (Wolthusen, 2009), the introduction of the newly emerging technologies in the cloud creates new difficulties for the digital forensics investigator but did not focus on the applicability of existing laws.

## 6.3 Sources of evidences

The nature of forensic evidences in the cloud environment is highly delicate and dynamic (Taylor, Haggerty, & D. Gresty, 2010). For instance, if an individual accesses an application through the cloud computing system, data written to the operating system will be stored in the virtual memory and hence, it will be lost if the user exits. Thus, these factors make the evidence extraction too complex. This section explains some of the sources of evidences that could be extracted by the investigator in the cloud environment. The possible locations of the cloud where the investigator could gather evidential data include network, hardware, host OS, hypervisor, and virtual machines (Rubsamen, Reich, Taherimonfared, Wlodarczyk, & Chunming Rong, 2013). Log data is a chief source of digital evidence that facilitates the reconstruction of the event series. Therefore, there is a need to maintain the log data with secure logging protocols. There are several secure logging protocols as listed in (Accorsi, 2009). It examined two main issues in the existing logging protocols. Most of the existing security logging protocols do not cover both the transmission phase and the storage phase of the log information. The second issue deals with the vulnerabilities of the protocols developed for specific phases over erroneous evidence. It concludes that none of the existing protocols completely satisfies the requirements for admissible evidence. It elucidates the various classes of secure logging protocols such as Syslog, Schneier/Kelsey, Ma/Tsudik, and Enc. Search.

### 6.4 Sources of evidence by logging

Log information is more important for forensic investigation in the cloud. Many researchers have discussed log analysis in the context of a cloud. The challenges involved in analyzing log information include decentralization, log volatility, multiple tiers and layers, log accessibility, absence of log information, and depending on the cloud service provider. The solution in (R. Marty, 2011) suggests a way to gather log information. It recommends concentrating on three information such as log time (e.g. Timestamp details, session ID), log purpose (reason) and log technique (application). Furthermore, it suggested syntax for logging as a key - value pair with three fields such as an object, action, and status. It also executed an application logging infrastructure at a SaaS company in which logging library was built that can be used in Django. Though, it suggests several techniques for collecting the log information, it does not address some important forensically sound data in IaaS and PaaS service models such as logging network usage, file metadata, and process usage.

## 6.5 Isolating a cloud instance

The isolation of the cloud instance prevents contamination of evidence and is a challenging task as it locates multiple instances in a node. Moving suspected instance from one node to another may result in loss of evidence. Therefore, (Delport & Olivier, 2012) suggested moving other instances in the same node. The following techniques are used for isolating the instances.

- Instance relocation
- Server farming
- Failover
- Address relocation
- Man-in-the-Middle (MITM)

## 6.6 Cloud Forensics Countermeasures

Though cloud forensic approaches are still relatively less effective, some of the approaches have contributed some progress. Technologies and approaches are discussed in (Lei & Cui, 2013). Service recreates the cloud service requests, renovates unauthorized access, and searches for potential evidence. Data migration is the key technology of the cloud storage system. The cloud service providers offer the service that migrates the manual data to cloud-based data backup and retrieval. The forensics process samples monitoring of data migration to gather possible evidence. Presenting the evidence gathered from a cloud-computing environment in a court of law is challenging due to the complexity of the data and also it is difficult to maintain the data integrity. The forensic examiner can use pattern matching and some other statistical analysis tools to ensure the integrity of the data and can trace the actions using the data fusion technology.

### 6.6.1 Identification of evidence

One of the major challenges faced with the data stored in cloud services is to identify the exact location of forensic-rich data. It requires more research for identifying and isolating the forensic evidences in the cyber domain, especially in cloud computing. In such cases, mobile devices introduce more difficulty to this cyber domain (Zatyko & Bay, 2012). In the mobile cloud environment, in most cases, the source of evidence on the client device is a web browser. Hence, information collected from a browser is taken into account in forensic examinations (D. Birk, 2011). Moreover, (Thorpe & Ray, 2012) suggested that locate the digital evidences in browsing history caches and this add difficulties in the collection, collation, and verification. Delay in accessing data as a result of jurisdictional issues could affect the investigation procedure. This kind of delay could result in data loss, and modification of data prior access is granted (Taylor, Haggerty, & D. Gresty, 2010). Hence, timely identification and preservation of evidence are essential.

### 6.6.2 Preservation of evidence

Data residing in the cloud can move from one location to another and disappears quickly, and collecting this data in a forensic manner is difficult. There is an urgent requirement for developing standard forensic procedures and set of tools to gather information from the cloud. Consider the volatile nature of data while collecting them from cloud sources (Ruan, Carthy, & T. Kechadi M. Crosbie, 2011). The research work in (Dykstra & Sherman, 2012)examines the technique of securing forensic-rich data from a remote infrastructure-as-a-service (IaaS) computer accessed through Amazon EC2. This examination depends on the installation of Encase Servlet or an FTK agent. There is an alternative insisting the investigator to request the information from the cloud service provider, but there is no guarantee in the words of service providers. Moreover, the service provider may not respond in time due to legal issues. The use of Encase Servlet or FTK agent does not make sense if the identified evidence resides in the cloud storage service because the user can access the data stored in the account and not in the storage environment.

### 6.6.3 Analysis of evidence

An analyst overwhelms with a large amount of data while dealing with the digital data and also turned out to be difficult to manage. Therefore, to make the data management process easier, it is necessary to filter the data related to the investigation and disregard other data. The process of classifying the relevant and non-relevant data to the investigation is difficult. Data analysis consumes more time and resources due to the examination of a large number of devices. Several organizations encrypt the data before uploading it to a cloud service (Taylor, Haggerty, & D. Gresty, 2010). It prevents the analyst from analyzing the data due to the unavailability of the decryption key. Forensic analyst should be aware of the basic procedures of evidence examination and analysis on the latest cloud computing systems (Taylor, J, & Lamb, 2011).

Hash analysis is a frequently used method to reduce data and refine the investigation-related files. A massive volume of distributed data in the cloud platform acts as a barrier to the use of signature detection methods and suggests applying the hashing process in a distributed manner across cloud platforms (Hegarty, Merabti, Shi, & Askwith, 2011). The absence of operating system metadata such as log files, unused space, and temporary files adds more difficulty in a conventional analytical process (Grispos, Storer, & Glisson).

### 6.6.4 Reporting and presentation of evidence

Digital evidences exist in a logical and physical context. Data is usually stored on the physical media that can be inferred through software (Reilly, Wren, & Berry, 2011). The investigation process mainly depends on the information residing on a mobile device (Taylor, J, & Lamb, 2011) . In contrast, (Bursztein, Cassidian, & Martin, 2011) declared that the analysis of the client's system may not be sufficient and hence, it is recommended to analyze the information stored in the cloud environment by the corresponding user. The presentation of evidences involves in collecting the sources from various contexts. It is also important to be aware of the forensic implications of gathering information from a cloud storage service to facilitate evidence to be reported and presented.

**6.7 Information provenance in cloud**

Provenance explains the history or origin of the data. Secure provenance could assist the investigator in getting the details of the owner of the data at a specific time, and the person who accessed the data including the time factor. However, even now, data provenance remains as undeveloped part in cloud computing. It is vital to secure the provenance information as leaking this information may result in the lack of confidentiality and user privacy (Lu, Lin, Liang, & Sherman, 2010). Secure provenance in cloud forensics plays an important role in preservation phase that can offer a sequential access history of forensic data and the way of examination. Some research works exploited the concept of provenance to cloud forensics (Dykstra & A. Sherman, Understanding issues in cloud forensics: Two hypothetical case studies, 2011) (D. Birk, 2011) .

**7.0 Forensic investigations in mobile cloud storage services**

Now-a-days, cloud storage services attract several people due to the easy accessibility. Cloud users exploit cloud storage services to manage and access documents from anywhere at any time. Malicious users can easily abuse cloud storage services that initiate forensic investigation procedure in cloud storage devices. The forensic examiners consider that there is a possibility to examine the source of evidences in cloud storage services only in PCs. However, it does not ensure the forensic examiner complete information. Cloud computing posits several difficulties to conventional digital forensic processes, as cloud servers are located in a jurisdiction i.e. different from the investigating suspect.

Most of the research work suggests solutions for issues in the server forensic analysis, such as logging and remote extraction of data (R. Marty, 2011) (Dykstra & Sherman, 2012) .

Recently, a very few research work focus on the forensic perspective on cloud storage products that deals with the mobile device forensic analysis (Chung, Park, Lee, & Kang, 2012) (Quick & Choo, Dropbox Analysis , 2013) (Quick & K.-K.R. Choo, Digital droplets: Microsoft SkyDrive forensic data remnants, 2013). It attempts to find out the presence of cloud storage and to extract the data relevant to forensic investigation. Recently developed forensic tools enable the investigator to investigate the interactions of the Smartphones with storage service providers, without directly accessing the data in the cloud service. However, there is no evidence for proving a relation between the data residing in the cloud and data retained in the Smartphone after an interaction. Table 4 reveals possible methods to collect evidences from the cloud storage service.

| Evidential artifacts in cloud storage services | |
|---|---|
| **Mobile devices** | **Cloud environment** |
| Sync metadata | Administrative metadata |
| Cached files and browser artifacts | Stored files |
| File management and encryption metadata | File management and encryption metadata |
| Cloud service and authentication data | Log and authentication data |

**Table 4: Evidential artifacts in cloud storage services**

**7.1 Forensic methodologies**

A forensic investigation model in (Chung, Park, Lee, & Kang, 2012) explained some vital components of the investigation. It described unknown forensic techniques of cloud storage services for four different operating systems such as Windows, Mac, iOS, and Android formerly. Moreover, it recommended a procedure to gather the remnants from PCs and Smartphones accessing cloud storage services. The artifacts created as a result of accessing cloud services are forensic-rich in nature. A wide range of artifacts can be recovered from the user profile by searching into logs, stored files, and databases. It focused on four commonly used cloud storage services such as Amazon S3, Dropbox, Google Docs, and Evernote. It examines the artifacts left on the user PCs of Mac and Windows computers. Cloud storage services can be accessed and examined the artifacts created by two popular web browsers: Internet Explorer and Firefox. Though there are valuable artifacts left behind while accessing web browser, most of the data resides in the temporary Internet files and browsing history. The information recovered from these artifacts indicates the accessed sites along with time. Dropbox and Evernote facilitate synchronization between the data residing in the cloud and PC hard drive. It assists the investigator to extract the actual files and corresponding metadata from the user's device. It performed a case study demonstrating the file synchronization in which it simply traced-out the file names instead of creating a hash value of the files to ensure the integrity of the file at every stage of the investigation.

Papers (Quick & Choo, Dropbox Analysis , 2013) and (Quick & K.-K.R. Choo, Digital droplets: Microsoft SkyDrive forensic data remnants, 2013) suggested a comprehensive analysis of recovering possible information from memory and permanent storage of the Windows PCs and Apple iPhone while accessing Dropbox and Skydrive through the browser or user applications. Similarly, analysis was conducted for Amazon Cloud Drive (Hale, 2013). On contrast, acquisition procedures and tools of file content and cloud object's metadata deserve a larger degree of intensifying. (D. Quick & Choo, 2013) Collected the maximum number of files from the user account of Dropbox, Microsoft SkyDrive, and Google Drive. The investigation lacked appropriate forensic software tools for collecting data at the primary level. Therefore, use of an internet browser or the client application could manage the consequences. These are only general purpose tools that are not forensically sound. Hence, recovering of valuable information like historical versions of life is impossible with these tools. The outcome of (D. Quick & Choo, 2013) stated that there were no changes in the contents even after downloading a file though only few timestamps preserved.

Cloud Data Imager Library is a mediation layer that facilitates the investigator a read-only access to the contents of the files. The metadata of particular remote folders and assists access to Dropbox, Microsoft SkyDrive, and Google Drive storage facilities (Corrado Federici, 2014). Moreover, it solves the issues regarding user credentials, access and refresh tokens. There is a possibility to take a snapshot of a system using EBS Boot volumes in Amazon cloud services. However, it is not suitable for cloud storage services as this process is out of the user's control (Reese, 2010). The authors of (Clark, 2011) analyzed picture Exif metadata remnants in Microsoft SkyDrive, Flickr, and Windows Azure and discovered that private information exists in several publicly shared pictures can have forensic-rich information like global positioning information.

## 7.2 Forensic Analysis in Dropbox

Dropbox is the most widely and frequently used file hosting service throughout the world. It allows users to store and share files and folders through web browser or client software. Dropbox encrypts the files and saves them in Amazon S3. The user can access the Dropbox account using iPhone through either inbuilt browser or the Dropbox App (Quick & Choo, Dropbox Analysis , 2013). This section describes the analysis of Dropbox accessed in Apple iPhone 3G. Access the Dropbox account using the application (Dropbox App) installed on the iPhone and view the file stored in the account.

### 7.2.1 Identification and collection

It identifies the files that are in the logical extraction. It includes the output of (.XRY) software with PDF reports and files uploaded using (.XRY).

### 7.2.2 Preservation

The extracted (.XRY) files are logical files. Therefore, these files are preserved in X-Ways Evidence File Container (ctr) format. It uses MD5 hash values to maintain the forensic data integrity.

### 7.2.3 Analysis

It examines each of the extracted files that has forensic value using forensic analysis tools including X-Ways Forensic version 16.5 and analyzes the content of the Apple plist files extracted from (.XRY) using Guidance Software Encase version 6.19.4.PList Explorer v1.0. SQLite Database Browser 2.0 analyzes the SQLite files. The examiner can determine the username with which the suspect accessed the Dropbox account in the ('com.getdropbox.Dropbox.plist') file. It does not extract the text from Enron files and locate the password from any of the extracted files.

## 7.3 Forensic Analysis in SkyDrive

The Apple iPhone 3G can access SkyDrive cloud storage in two ways: inbuilt Safari browser and Microsoft SkyDrive App. Some of the hardware and software solutions that support, forensic practitioners in analysis of iPhone devices are Microsystemation (.XRY), Cellebrite UFED, and Radio Tactics Aceso. This section reveals the kinds of information that could be obtained from iPhone accessing the SkyDrive. Moreover, this section does not focus on analyzing the portable device accessed through a browser instead it focuses on SkyDrive App. Consider an iPhone that hasn't accessed or used SkyDrive App previously. The Microsystemation (.XRY 6.2.1) extracts the logical image of the iPhone content, except video and audio files prior installing SkyDrive App. The logical extraction confirms that there were no data associated with the use of SkyDrive. Accessing the application with a user account to view the stored files (if any) in the SkyDrive account to perform the (.XRY) logical extraction.

### 7.3.1 Identification and collection

In this case, files needed to be analyzed are ( .XRY) extract files and the outcome of (.XRY) software. Collect the files extracted at each stage such as base, browser, and application.

### 7.3.2 Preservation

As per principles of forensic investigation phases, it preserves the (.XRY) extracted files, output files, and reports after collecting them. As the extracted files are logical files, these were stored in the Encase Logical format (L01), and X-Ways Evidence File Container format (ctr). MD5 ensures the forensic data integrity using hash values.

### 7.3.3 Analysis

Forensic tools like X-Ways Forensic 16.5 and Encase 6.19.4 analyzes the forensic logical files. It examines the contents of the Apple plist files extracted from (.XRY) files using PList Explorer v1.0. Analysis of files extracted by Base-XRY does not disclose the presence of data related to Enron sample test data and SkyDrive App includes the website's URL. The analysis of the browser using the (. XRY) logical extraction method could not locate the username, but locates the client ID number in the corresponding URL (skydrive.live.com) in (History.plist) file. Moreover, it also locates the SkyDrive OwnerID in the (Cookies.binarycookies) file. Analysis of SkyDrive App i.e. third XRY logical extraction could find-out the username used to access SkyDrive account. This level of extraction could not locate the text from the Enron files and passwords in any of the (.XRY) extracted files.

### 9.0 Opportunities in mobile cloud forensics

Mobile cloud-based applications have attracted users with convenient applications that offer significant information improving productivity. The current forensic tools deployed in the mobile cloud environment are not 100% forensically sound. The opportunities of mobile cloud forensics in the future include the development of new techniques and algorithms for acquiring more data from the mobile device and the cloud. Moreover, there is an opportunity to include more forensic tools and make more benchmarks that incorporate the features of several handheld devices. Future research on mobile cloud forensics will focus on developing standard forensic analysis of mobile devices in various cloud services. Additionally, mobile cloud forensics provides more job opportunities in all the application fields of mobile clouds such as network operators, application developers, vendors, and service providers apart from the digital forensic investigators. Apart from forensics techniques, there are some anti-forensic techniques that facilitate to modify, and delete, securely and selectively, the digital evidence on the Smartphone without using cryptographic mechanisms (Albano, Castiglione, Cattaneo, & Santis, 2011).

**Conclusion**

Mobile cloud computing offers potential applications for mobile users as it merges the advantages of both mobile and cloud computing. The popularity and offers of mobile cloud computing motivated the illegal activities around the globe. Today's forensic tools, methodologies, and procedures are facing several challenges to solve the investigation issues in mobile cloud computing. This paper presents a comprehensive review and detailed analysis of forensic investigation in a mobile device, cloud and mobile cloud applications. It addressed the limitations and the impact of digital forensics in mobile cloud applications. There are several constraints in conducting digital investigations in the mobile cloud environment. It discussed the basic steps of mobile device and cloud forensics such as identification and preservation, collection, examination and analysis, and presentation. It discussed several techniques and tools in obtaining and analyzing forensic rich data from a mobile device and related sources. Similarly, it also analyzed the forensic approaches and tools in acquiring forensic rich data from cloud services and storage. Heterogeneity in a device, operating system, file format, applications, cloud platforms and storage, distributed architecture and dynamic environment are the potential factors which hamper forensic investigation in mobile cloud applications. In order to overcome these factors, extensive forensic research is inevitable to cope with the technology advancement in the mobile cloud platform. In future, mobile cloud forensic should focus on standards, cooperation, scalability, validation and inter-operability among the tools and techniques.

# Bibliography

Accorsi, R. (2009). Safekeeping Digital Evidence with Secure Logging Protocols: State of the Art and Challenges. *IEEE 5th International Conference on IT Security Incident Management and IT Forensics (IMF)*, (pp. 94-110).

Albano, P., Castiglione, A., Cattaneo, G., & Santis, A. D. (2011). A novel anti-forensics techniques for the Android OS. *IEEE International conference on broadband and wireless computing, communication, and applications*, (pp. 380-385).

Al-Saleh, M. I., & Forihat, a. Y. (2013). Skype Forensics in Android Devices. *International Journal of Computer Applications,*, 38-44.

Aminnezhad, A., Dehghantanha, A., Abdullah, M. T., & ohsen Damshenas. (2013). Cloud Forensics Issues and Opportunities . *International Journal of Information Processing and Management (IJIPM),*, 76-85.

Andriotis, P., Oikonomou, G., & Tryfonas, T. (2012). Forensic analysis of wireless networking evidence of android smartphones. *IEEE International Workshop on Information Forensics and Security (WIFS)*, (pp. 109-114).

Bursztein, E., Cassidian, I., & Martin, M. (2011). Doing Forensics in the Cloud Age Owade: Beyond Files Recovery Forensic. *Black Hat*.

Buyyaa, R., Yeoa, C. S., Venugopala, S., Broberg, J., & Brandic, I. (2009). Cloud Computing and Emerging IT Platform:Vision, Hype, and reality for delivering computing as the 5th utility. *Elsevier Transaction on Future generation Computer System*, 599-616.

Casadei, F., Savoldi, A., & Gubian, P. (2006). Forensics and SIM cards: an Overview. *International Journal of Digital Evidence*.

Chung, H., Park, J., Lee, S., & Kang, C. (2012). Digital forensic investigation of cloud storage services. *Digital Investigation*, 81-95.

Clark, P. (2011). Digital Forensics Tool Testing–Image Metadata in the Cloud. *Department of Computer Science and Media Technology, Gjovik University College,* , (pp. 1-54).

Corrado Federici. (2014). Cloud Data Imager: A unified answer to remote acquisition of cloud storage areas. *Elsevier transaction on Digital Investigation*, 30-42.

counton2. (n.d.). *counton2*. Retrieved from www.counton2.com/story/26564587/mobile-cloud-market-worth-4690billion-by-2019-new-report-by-marketsandmarkets .

D. Birk. (2011). Technical challenges of forensic investigations in cloud computing environments. *IEEE 6th international workshop on Systematic Approaches to Digital Forensic Engineering (SADFE)*, (pp. 1-10).

D. Quick, & Choo, K. (2013). Forensic collection of cloud storage data: does the act of collection result in changes to the data or its metadata? *Digital Investigation*, 266-277.

Daryabar, F., Dehghantanha, A., Udzir, N. I., & Mohd, N. F. (2013). A Survey about Impacts of Cloud Computing on Digital Forensics. *International Journal of Cyber-Security and Digital Forensics (IJCSDF), The Society of Digital Information and Wireless Communications*, 77-94.

Delport, M. K., & Olivier, M. S. (2012). Isolating a cloud instance for a digital forensic investigation. *Springer, Advances in Digital Forensics VIII, International Federation for Information Processing (IFIP) Advances in Information and Communication Technology*, 187-200.

Dinh, H. T., Lee, C., Niyato, D., & Wang, P. (2013). A Survey of Mobile Cloud Computing. *Wireless Communication and Mobile Computing*, (pp. 1-38).

Dykstra, & Sherman, A. (2012). Acquiring forensic evidence from infrastructure-as-a-service cloud computing: exploring and evaluating tools, trust, and techniques. *Digital Investigation,*, 90-98.

Dykstra, J., & A. Sherman. (2011). Understanding issues in cloud forensics: Two hypothetical case studies. *Journal of Network Forensics*, 19-31.

Dykstra, J., & A.T. Sherman. (2012). Acquiring forensic evidence from infrastructure-as-a-service cloud computing: exploring and evaluating tools, trust, and techniques. *Digital Investigation*, 90-98.

Dykstra, J., & Sherman, A. (2012). Acquiring forensic evidence from infrastructure-as-a-service cloud computing: exploring and evaluating tools, trust, and techniques. *Digital Investigation*, 90-98.

Fernando, N., Loke, S. W., & Rahayu, W. (2013). Mobile cloud computing: A survey. *ELSEVIER transaction on Future Generation Computer Systems, Vol. 29*, 84-106.

Gao, F., & Zhang, Y. (2013). Analysis of WeChat on iPhone. *2nd International Symposium on Computer, Communication, Control, and Automation (3CA),*, (pp. 278-281).

Goasduff, L., & Pettey, C. (2012). *Gartner says worldwide Smartphone sales soared in fourth quarter of 2011 with 47 percent growth.* Gartner, Newsroom, Announcements, press release.

Grispos, G., Glisson, W. B., & Storer, T. (2013). Using Smartphones as a Proxy for Forensic Evidence contained in Cloud Storage Services. *46th Hawaii International Conference on System Sciences,* , (pp. 1-10).

Grispos, G., Storer, T., & Glisson, W. (n.d.). Calm before the storm: The challenges of cloud computing in digital forensics. *International Journal of Digital Crime and Forensics (IJDCF)*, 2012.

Grover, J. (2013). Android forensics: automated data collection and reporting from a mobile device. *Digital Investigation, proceedings of 13th annual digital forensics research conference,*, (pp. 12-20).

Guido, M., Ondricek, J., Grover, J., Wilburn, D., Nguyen, T., & Hunt, A. (2013). Automated identification of installed malicious android applications. *The proceedings 13th annual digital forensics research conference*, (pp. 96-104).

Hale, J. (2013). Amazon cloud drive forensic analysis. *Digital Investigation*, 259-265.

Hegarty, R., Merabti, M., Shi, Q., & Askwith, B. (2011). Forensic Analysis of Distributed Service Oriented Computing Platforms. *12th Annual Post-Graduate Symposium on the Convergence of Telecommunications, Networking and Broadcasting.*

IDC. (n.d.). *IDC*. Retrieved from www.idc.com/prodserv/smartphone-os-market-share.jsp .

Kent, ]. K., Chevalier, S., Grance, T., & Dang, H. (2006). Guide to integrating forensic techniques into incident response. *Department of Commerce, National Institute of Standards and Technology, Computer Security.*

Khan, A. u., Othman, M., Madani, S. A., & Khan, S. U. (2013). A Survey of Mobile Cloud Computing Applicaiton Models. *IEEE communications society* , (pp. 393-413).

Lee, J., & Dowon Hong. (2011). Pervasive Forensic Analysis based on Mobile Cloud Computing. *IEEE computer society, 3rd International Conference on Multimedia Information Networking and Security*, (pp. 572-576).

Lei, Y., & Cui, Y. (2013). Research on Live Forensics in Cloud Environment. *2nd International Symposium on Computer, Communication, Control, and Automation (3CA)*, (pp. 231-234).

Lessard, J., & Kessler, G. C. (2009). Android forensics: Simplifying cell phone examinations. *Small Scale Digital Device Forensics Journal*.

Levinson, A., Stackpole, B., & Johnson, D. (2011). Third party application forensics on apple mobile devices. *IEEE proceedings of 44th Hawaii International Conference on System Sciences*, (pp. 1-9).

Liu, Q., Jain, X., Hu, J., Zhao, H., & Zhang, S. (2009). An optimized solution for mobile environment using mobile cloud computing. *5th International Conference on Wilreless Communications, Networking and Mobile Computing*, (pp. 1-5).

Lovell, R. (2011). White paper: Introduction to cloud computing. *ThinkGrid.*

Lu, R., Lin, X., Liang, X., & Sherman, X. (2010). Secure Provenance: The Essential of Bread and Butter of Data Forensics in Cloud Computing. *ACM, 2010.*

Me, G. (2008). Internal forensic acquisition for mobile equipment". *IEEE proceedings of the International Symposium on Parallel and Distributed Processing*, (pp. 1-7).

Meng, F., Wu, S., Yang, J., & Yu, G. (2009). Research of an E-mail Forensic and Analysis System Based on Visualization. *IEEE 2nd Asia-Pacific Conference on Computational Intelligence and Industrial Applications*, (pp. 281-284).

Meulen, R. v., & J. Rivera. (2013). *Gartner says worldwide mobile phone sales declined 1.7 percent in 2012.* Gartner.

Michael Armbrust, A. F., Lee, G., Patterson, D. A., Rabkin, A., Stoica, I., & Zaharia, M. (2009). Above the Clouds: A Berkeley View of Cloud Computing. *Technical Report, UC Berkeley Reliable Adaptive Distributed Systems Laboratory*, (pp. 1-23).

Mutawa, N. A., Baggili, I., & Marrington, A. (2012). Forensic analysis of social networking applications on mobile devices. *Elsevier transaction on Digital Investigation,*, 24-33.

NetQin. (2011). *Growing conflict between IT and mobility*. Retrieved from www.netqin.com/en/security/newsinfo_4094_3.html

Nielsen. (2012). *State of the appnation —a year of change and growth in U.S. Smartphones.* Nielsen.

Quick, D., & Alzaabi, M. (2011). Forensic analysis of the android file system yaffs2. *Proceedings of the 9th Australian digital forensics conference*, (pp. 100-109).

Quick, D., & Choo, K. (2013). Dropbox Analysis . *Digital Investigation*, 1378-1394.

Quick, D., & Choo, K. (2013). Dropbox analysis: Data remnants on user machines. *Digital Investigation*, 3-18.

Quick, D., & K.-K.R. Choo. (2013). Digital droplets: Microsoft SkyDrive forensic data remnants. *Future Generation Computer Systems,*, 1378-1394.

R. Marty. (2011). Cloud application logging for forensics,. *ACM Symposium on Applied Computing*, (pp. 178-184).

Raghav, S., & Saxena, A. K. (2009). Mobile Forensics: Guidelines and Challenges in Data Preservation and Acquisition. *Proceedings of Student Conference on Research and Development (SCOReD)*, (pp. 16-18).

Reese, G. (2010). *Cloud Forensics Using EBS Boot Volumes*. Retrieved from www.Oreilly.com

Reilly, D., Wren, C., & Berry, T. (2011). Cloud Computing: Pros and Cons for Computer Forensic Investigations. *International Journal Multimedia and Image Processing (IJMIP), Infonomics Society*, 26-34.

Research, D. (2012). *The impact of mobile devices on information security: A survey of it professionals.*

Research:, A. (2009). *Mobile Cloud Computing - Next Generation Browsers, Widgets, SIM, Network-as-a-Service, and Platform-as-a-Service, 3Q*. Retrieved from http://www.abiresearch.com/research/1003385.

Ruan, K., Carthy, J., & T. Kechadi M. Crosbie. (2011). Cloud forensics An overview. *7th IFIP International Conference on Digital Forensics.*

Ruan, K., Carthy, J., Kechadi, T., & Baggili, I. (2013). Cloud forensics definitions and critical criteria for cloud forensic capability: an overview of survey results. *Elsevier journal on Digital Investigation,*, 34-43.

Rubsamen, T., Reich, C., Taherimonfared, A., Wlodarczyk, T., & Chunming Rong. (2013). Evidence for Accountable Cloud Computing Services. *Pre-Proceedings of International Workshop on Trustworthiness, Accountability and Forensics in the Cloud (TAFFC).*

Sibiya, G., Venter, H. S., & Thomas Fogwill. (2012). Digital Forensic Framework for a Cloud Environment. *IST-Africa Conference Proceedings on International Information Management Corporation (IIMC)*, (pp. 1-8).

Simson L. Garfinkel. (2010). Digital forensics research: The next 10 years. *Elsevier, Digital Investigation,*, 64-73.

Son, N., Lee, Y., D. Kim, J. J., Lee, S., & Lee, K. (2010). A study of user data integrity during acquisition of android devices Digital Investigation,. *DFRWS conference 13th annual digital forensics research conference*, (pp. 1-12).

Sylve, J., Case, A., Marziale, L., & Richard, G. (2012). Acquisition and analysis of volatile memory from android devices. *Digital Investigation, .*

Taylor, M. H., J, G. D., & Lamb, D. (2011). Forensic Investigation of Cloud Computing Systems. *Network Security,*, 4-10.

Taylor, M., Haggerty, J., & D. Gresty, R. H. (2010). Digital evidence in cloud computing systems. *Digital Investigation, computer law and security review*, (pp. 304-308).

Thorpe, S., & Ray, I. (2012). Cloud Log Forensics Metadata Analysis. *IEEE 36th Annual, Computer Software and Applications Conference Workshops (COMPSACW)*, (pp. 194-199).

Turnbull, B., & & Slay, J. (2008). WiFi network signals as a source of digital evidence: Wireless network forensics. *IEEE 3rd international conference on Availability, Reliability and Security*, (pp. 1355-1360).

Vidas, T., Zhang, C., & Christin, N. (2011). Toward a general collection methodology for android devices. *Digital Investigation*.

Wang, Y., Chen, I.-R., & Wang, a. D.-C. (2014). A Survey of Mobile Cloud Computing Applications: Perspectives and Challenges. *Springer transaction on wireless Personal Communications.*

Wolthusen, S. D. (2009). Overcast: Forensic Discovery in Cloud Environments. *5th International Conference on IT Security Incident Management and IT Forensics, IEEE Computer Society*, (pp. 3-9).

Yates, M. (2010). Practical Investigations of Digital Forensics Tools for Mobile Devices. *ACM conference on Information Security Curriculum Development Conference*, (pp. 156- 162).

Zatyko, K., & Bay, J. (2012). *The Digital Forensics Cyber Exchange Principle.* Forensic Magazine.

Zdziarski, J. (2014). Identifying back doors, attack points, and surveillance mechanisms in iOS devices. *Elsevier transaction on Digital Investigation,* , 3-19.

Zdziarski, J., & Media, O. (2008). iPhone forensics: Recovering evidence, personal data, and corporate assets. *Digital Forensics Solutions*.

Zhu, M. (2011). *Mobile Cloud Computing: Implications to Smartphone Forensic Procedures and Methodologies" .* AUT University Theses and Dissertations.